\documentclass{article}
\usepackage{hyperref}
\usepackage{url}
\usepackage{makecell}
\usepackage{multirow}
\usepackage[super]{nth}

\usepackage{tikz, pgfplots, pgfplotstable, filecontents, xcolor}
\usepackage{spconf,amsmath,graphicx,amssymb}

\title{Emotional Speech Synthesis with Rich and Granularized Control}

\name{Se-Yun Um$^1$, Sangshin Oh$^1$, Kyungguen Byun$^1$, Inseon Jang$^2$, ChungHyun Ahn$^2$, Hong-Goo Kang$^1$ }
\address{$^1$Yonsei University, Department of Electrical and Electronic Engineering, Seoul, South Korea \\
$^2$Electronics and Telecommunications Research Institution, Daejeon, South Korea\\}

\begin{document} 
\ninept
\maketitle 

\begin{abstract}
This paper proposes an effective emotion control method for an end-to-end text-to-speech (TTS) system.
To flexibly control the distinct characteristic of a target emotion category, it is essential to determine embedding vectors representing the TTS input. We introduce an inter-to-intra emotional distance ratio algorithm to the embedding vectors that can minimize the distance to the target emotion category while maximizing its distance to the other emotion categories. To further enhance the expressiveness of a target speech, we also introduce an effective interpolation technique that enables the intensity of a target emotion to be gradually changed to that of neutral speech.
Subjective evaluation results in terms of emotional expressiveness and controllability show the superiority of the proposed algorithm to the conventional methods.

\end{abstract}
\section{Introduction}
\label{sec:intro}
The objective of a text-to-speech (TTS) system is to synthesize human-like speech signals such that linguistic and paralinguistic information can be conveyed clearly. Thanks to recent advances in deep learning-based TTS technologies, understanding the contextual meaning of text from synthesized speech is no longer an issue \cite{zen2013statistical, qian2014training, x39,oord2016wavenet,sotelo2017char2wav,arik2017deep,gibiansky2017deep}. However, it remains difficult to express paralinguistic information such as emotion through synthesized speech.

In deep learning-based end-to-end TTS systems (e.g., Tacotron), the expressiveness of synthesized speech is controlled by conditioning additional embedding vectors that implicitly provide prosody-related latent features typically generated by another neural network \cite{DBLP:journals/corr/abs-1711-05447, wang2018style, zhang2019learning,wang2017uncovering,bian2019multi, lee2019robust}. 
However, because those systems only mimicked a generic speaking style of reference audio, it was difficult to assign user-defined emotion types to synthesized speech. Lee et al. \cite{DBLP:journals/corr/abs-1711-05447} directly provided an emotion label to the decoder of the Tacotron system such that it was concatenated with the output of pre-net. Although the method showed a feasibility of presenting emotion to synthesized speech, it was also challenging to control emotional expressiveness flexibly.

A better approach was to learn a representative embedding style of each emotion by analyzing the distribution of an emotion labeled database.
Because the embedding vectors obtained from the same emotion category contain similar prosodic information, they have a tendency to form a cluster. Kwon \textit{et al.} \cite{kwon2019effective} effectively generated emotional speech using representative condition vectors obtained by averaging the style embedding vectors belonging to each emotion category.
However, we found out that the approach could not clearly present the distinctive characteristics of each emotion.
Because the distribution of style embedding vectors in the embedding domain were highly dispersed, their mean was not a good representative vector. 

The main objective of this work is to develop a TTS system that not only generates each emotional speech clearly but also flexibly controls the intensity of emotion representation.  
To achieve this, we focus on (i) how to determine representative embedding vectors for each emotion, and (ii) how to control flexibly the intensity of emotion by adjusting representative vectors. 
At first, we propose an inter-to-intra distance ratio algorithm that utilizes the ratio of the distance between intra-cluster embedding vectors and inter-cluster ones. Unlike the previous approach that only considers a target emotion category, the proposed approach considers other emotion categories simultaneously. 
As the representative embedding vector reflects the characteristic of each emotion category well, the emotional clarity is further enhanced. In addition, we propose an effective interpolation technique that controls emotion intensity by gradually changing the prosody of the target emotion to that of neutral speech.

We investigated the effectiveness of the proposed method by conducting subjective evaluation tasks.
The experimental results show that our system successfully achieves better emotional expressiveness and controll-ability than conventional methods.

The rest of the paper is organized as follows.
Section \ref{sec:related_works} describes the related works and Section \ref{sec:proposed_method} explains the proposed algorithm. Experiments and conclusion are followed in Section \ref{sec:experiments} and \ref{sec:conclusion}, respectively.

\section{related works}
\label{sec:related_works}
The GST-Tacotron,which consists of an end-to-end TTS and a global style token module flexibly changes the style of synthesized speech to have characteristics similar to auxiliary input reference audio \cite{wang2018style}. 
It consists of the Tacotron network and a style encoder.
The Tacotron network estimates the mel-spectrogram of the output speech signal from input text using a sequence-to-sequence with attention mechanism \cite{chorowski2015attention, sutskever2014sequence, bahdanau2014neural, luong2015effective, chan2016listen, shen2018natural}. 
To train the style encoder, a mel-spectrogram of the reference audio is fed into the style encoder module to generate a fixed-length prosody embedding \cite{skerry2018towards}, then a multi-head attention method \cite{vaswani2017attention} is applied to present the similarity between prosody embedding and global style tokens (GST). Finally, style embeddings generated by the weighted sum of GSTs, as well as hidden features estimated by the text encoder are fed into the decoding module of the Tacotron network. 
The inference step is almost identical to the training step described above. It generates a mel-spectrogram using text input and style embedding obtained by either reference audio or selected by the emotion category, then the generated mel-spectrogram is fed into the WaveNet vocoder to synthesize a human-like emotional speech signal. 

\begin{figure}[t]
    \centering
    \includegraphics[width=0.8\columnwidth]{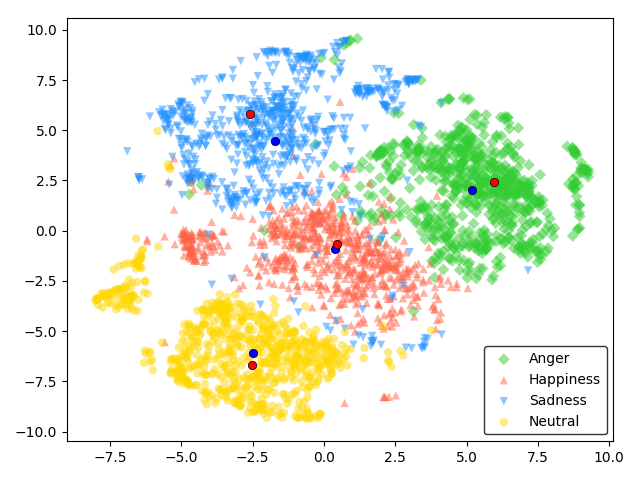}
    \caption{t-SNE plot of style embedding vectors' weight vectors for each emotion; anger, happiness, sadness and neutral. Representative points from mean-based approach are marked with blue and proposed points are red color.}
    \label{fig:t-sne}
\end{figure}

To generate emotional speech by controlling embedding vectors, however, the GST-Tacotron also requires a hand-crafted procedure that must identify which dimension of style embedding affects any attribution of speech \cite{wang2018style}. 
Kwon \textit{et al.} \cite{kwon2019effective} assumed the mean weights of a style embedding vector could be used to synthesize emotional speech as shown in Figure \ref{fig:t-sne}. In the training step, weights were first obtained from the pre-defined emotion category such as happiness, sadness, anger, and neutral. Because the weights obtained from the same emotion category could be clustered, the mean weights of each cluster were regarded as representative style embedding vectors. However, it is uncertain whether mean vectors should be the best choice to represent the characteristic of each emotion category. 

\textbf{Relationship to prior work}.
To determine the representative style embedding vector of each emotion to control the expressiveness of each emotion flexibly, we consider not only the weights' distribution of each emotion category but also the distances between each category. We also propose an effective interpolation technique to change the intensity of expressiveness smoothly by considering the vector space of each emotion cluster. 

\section{proposed model}
\label{sec:proposed_method}
It is natural to assume that the style embedding vectors having prosody of speech signals, which belong to the same emotion category, locates closely. 
Therefore, the weight vectors from the style token layer tend to form a cluster if they originate from the same emotion category. 
Figure \ref{fig:t-sne} shows an t-distributed stochastic neighbor embedding (t-SNE) plot \cite{maaten2008visualizing}visualizing 2-D representation from 40-dimensional style tokens' weight vectors, which form clusters depending on their emotion labels. 
To synthesize emotional speech from these clusters, a single style vector should be extracted, which we refer to as a representative weight.
In this section, we propose a method to determine effectively the representative weight vectors of each emotion category such that they clearly present unique emotion characteristics.
Then, we also propose an effective interpolation algorithm to control the intensity of emotional expressiveness flexibly, e.g. from weak emotion to strong emotion.

\subsection{Representative style vector}
\label{concept}
The representative style token weight vectors should satisfy two requirements. First, they should synthesize desired emotional speech without quality degradation.
Second, the representative weight vectors must be located within the boundary of each emotion category, otherwise the synthesis model fails to generate speech in the interpolation process when style token weights out of range are given to the model.

The simplest method to determine the representative weights $\mathbf{r}_e$ for emotion $e$ is averaging all the weights belonging to the corresponding emotion category as follows~\cite{8778667}:
\begin{equation}
    \mathbf{r}^{Mean}_e = \frac{1}{{N}_e} 
    \sum_{\mathbf{x}^{(i)} \in {X}_e } {\bf \mathbf{x}^{(i)}},
\end{equation}
where ${N}_e$, and $\bf \mathbf{x}^{(i)}$ are the number of weights and weight vector samples of the target emotion category, $e$, respectively.
However, because the mean-based method does not consider the distribution of weights such as variance, it cannot fully represent the information of the target emotion. It is also disadvantageous during the emotion interpolation process because the approach ignores the distribution of other emotion category. 

\label{subsubsec:I2I}
The key idea of the proposed method is to consider the distribution of the target and other emotion categories simultaneously. 
Specifically, the representative weight vectors, $r^{I2I}_e$, are determined as the ones that maximize the ratio of inter-category distance over intra-category distance (I2I).
\begin{align}
    \mathbf{r}^{I2I}_e 
    = \frac{1}{2}\arg \max_{\mathbf{r}}
    \frac {\mathbb{E}_{\mathbf{x} \in {X}_l}
          [ \| \mathbf{r} - \mathbf{x} \|_2 ]}
          {\mathbb{E}_{\mathbf{x} \in {X}_e}
          [ \| \mathbf{r} - \mathbf{x} \|_2 ]} \nonumber \\
          +
    \frac{1}{2}\arg \max_{\mathbf{r}}
    \frac {\mathbb{E}_{\mathbf{x} \in {X}_s}
          [ \| \mathbf{r} - \mathbf{x} \|_2 ]}
          {\mathbb{E}_{\mathbf{x} \in {X}_e}
          [ \| \mathbf{r} - \mathbf{x} \|_2 ]},
\end{align}
where ${X}_e$ denotes the weight vectors belonging to the target emotion category and ${X}_l$ and ${X}_s$ represent the weight vectors of the farthest and the closest emotion class, respectively. 
Consequently, the representative weight vectors are the ones that maximize the distances from the farthest and closest emotion categories while minimizing the distance within the target emotion category. Because the farthest emotion cluster has opposite characteristics, it can maximize its unique properties by selecting the farthest one. In addition, choosing the closest emotion makes it more distinguishable from others. 
As shown in Figure \ref{fig:t-sne}, representative style vectors determined by the proposed I2I method are located at greater distances from other style vectors than the conventional one.


\subsection{Emotion intensity control}
\begin{figure*}[t]
    \centering
    \includegraphics[width=0.8\textwidth]{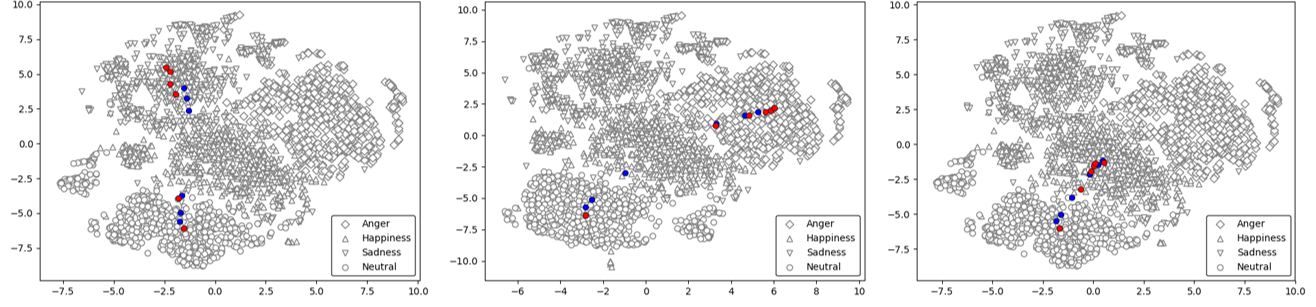}
    \caption{t-SNE plot of $\alpha$ calculated by linear interpolation (blue points) and proposed method (red points) where the target emotion is sadness (left), anger (middle) and happiness (right). }
    \label{fig:alpha}
\end{figure*} 

To control flexibly the intensity of emotion expressiveness, we propose an effective interpolation technique that gradually changes the weight values of emotion embedding vectors.
Because all the processing is done in the emotion embedding vector space, it is possible to synthesize signals having new types of emotions such as weak anger, gloomy sad, or strong happiness without altering other modules.

First, we introduce a simple solution called the linear interpolation method between two representative weight values. 

\begin{equation}
\label{lin_inter}
    \mathbf{r}^{Lin}_{e_1 : e_2} (\alpha)
    = \alpha \cdot \mathbf{r}_{e_1}
    + (1 - \alpha) \cdot \mathbf{r}_{e_2},
\end{equation}
where $\alpha \in [0, 1]$ is a scale factor to the first emotion category, and ${r}_{e_i}$ indicates the representative weight of each emotion category. However, because linearly interpolated representative weights always lie on the straight line between two representative weights, it is  often located at the outside of the emotion cluster region, which results in generating unnatural synthesized speech.
To change the intensity of emotion expressiveness smoothly without this artifact, we need to consider not only the representative weight vectors but also the distribution of the vectors in each emotion category. 
In addition, as the intensity of any emotion can be gradually changed from the neutral category, we set the neutral category as the reference point for a target emotion to be interpolated. 

In the second approach, called spread-aware I2I (SA-I2I) method, we artificially generate multiple weight vector samples to have a distribution of desired emotion intensity. Then, we find the representative weight vector using the I2I-based method described in Section~\ref{concept}.
The weight vector samples belonging to the interpolated category of scale $\alpha$, $X_{n : e} (\alpha)$, are obtained as follows: 
\begin{eqnarray}
    X_{n:e} (\alpha) = 
        \{  \frac{\mathbf{x} + \mathbf{y}}{2}
            \; | \;
            \mathbf{x} \in X'_{n \rightarrow e} (\alpha) , \,
            \mathbf{y} \in X'_{e \rightarrow n} (\alpha)
        \},\\
    X'_{n \rightarrow e} (\alpha)=
        \{  \alpha \mathbf{x} + (1 - \alpha) \mathbf{r}_{e} 
            \; | \; \mathbf{x} \in X_{n}
        \},  \nonumber \\      
    X'_{e \rightarrow n} (\alpha)=
        \{  (1 - \alpha) \mathbf{y} + \alpha \mathbf{r}_{n}
            \; | \; \mathbf{y} \in X_{e}
        \}, \nonumber       
\end{eqnarray}
where $X_n$ and $X_e$ are neutral and emotion clusters, and $\mathbf{r}_n$ and $\mathbf{r}_e$ are representative vectors of the neutral and target emotion category, respectively.

Because the intensity of each emotion varies more explicitly within the target emotion cluster region, it is reasonable to non-linearly determine the interpolation ratio, $\alpha$.
In addition, we empirically found that the change in emotion intensity was greater when the style token weight was close to the representative style token.

To find non-linear interpolation ratios that satisfy the conditions, we must find an anchor point that represents a boundary region of the neutral cluster and, then determine non-linear steps from the anchor point to the target emotion.

The anchor point, $\mathbf{b}_e$, is determined by the ratio of the function of the standard deviation of the weight vectors belonging to the neutral and target emotion category as follows.

\begin{eqnarray}
    \mathbf{b}_e = \frac{f(\sigma_{n})}{f(\sigma_{n}) + f(\sigma_{e})},\,
    {\sigma_p} = \frac{1}{M}\sum_{m=1}^{M} {\bf \mathbf{\sigma}_{p,m}}, 
\end{eqnarray}
where $\sigma_{n}$ and $\sigma_{e}$ are the average standard deviation of the weight vectors in the neutral and target emotion category. $\mathbf{\sigma_{p,m}}$ is the standard deviation of the $m$-th dimension of emotion $p$ which denotes the label of neutral or one of the emotion categories and $m$ is the number of vectors belonging to $p$. We adopted a square function, i.e. $f(\sigma)=\sigma^2$.
Then, the interpolation interval is evenly divided in exponential-scale from neutral to the representative weight vector of the target emotion.
The non-linear interpolation ratio, $\alpha_i$ is defined as follows: 
\begin{equation}
    \alpha_i = \log\{{e^{\mathbf{b}_e}  + \delta*(i-1)}\}, 1 \leq i \leq N,
\end{equation}
where $\delta$ is the non-linear step from the anchor point to the representative vector of the target emotion, and it is defined in the exponential domain as follows:
\begin{equation}
    \delta = \frac{e - e^{\mathbf{b}_e}}{N-1},
\end{equation}
where $N$ denotes a granularity level for controlling emotion intensity. $N$ can be an arbitrary integer but it was set to 4 for the experiments in this paper. 
The interpolation ratio of the representative vector of the target emotion is considered 1. Because the output of its approach is not a single representative weight vector but a distribution, we apply the I2I method proposed in Section \ref{subsubsec:I2I} to determine the representative weights of the interpolated cluster. Figure \ref{fig:alpha} depicts the estimated weight vectors to be used for synthesizing emotional speech. Compared to the linear-based approach, the distance between the representative values of each emotion is further increased and the spacing between the interpolated representative weight vectors is non-linear. This helps to enhance the expressiveness of synthesized speech compared to the conventional linear interpolation-based method.   


\begin{figure}[t] 
    \centering
    \includegraphics[width=1\columnwidth]{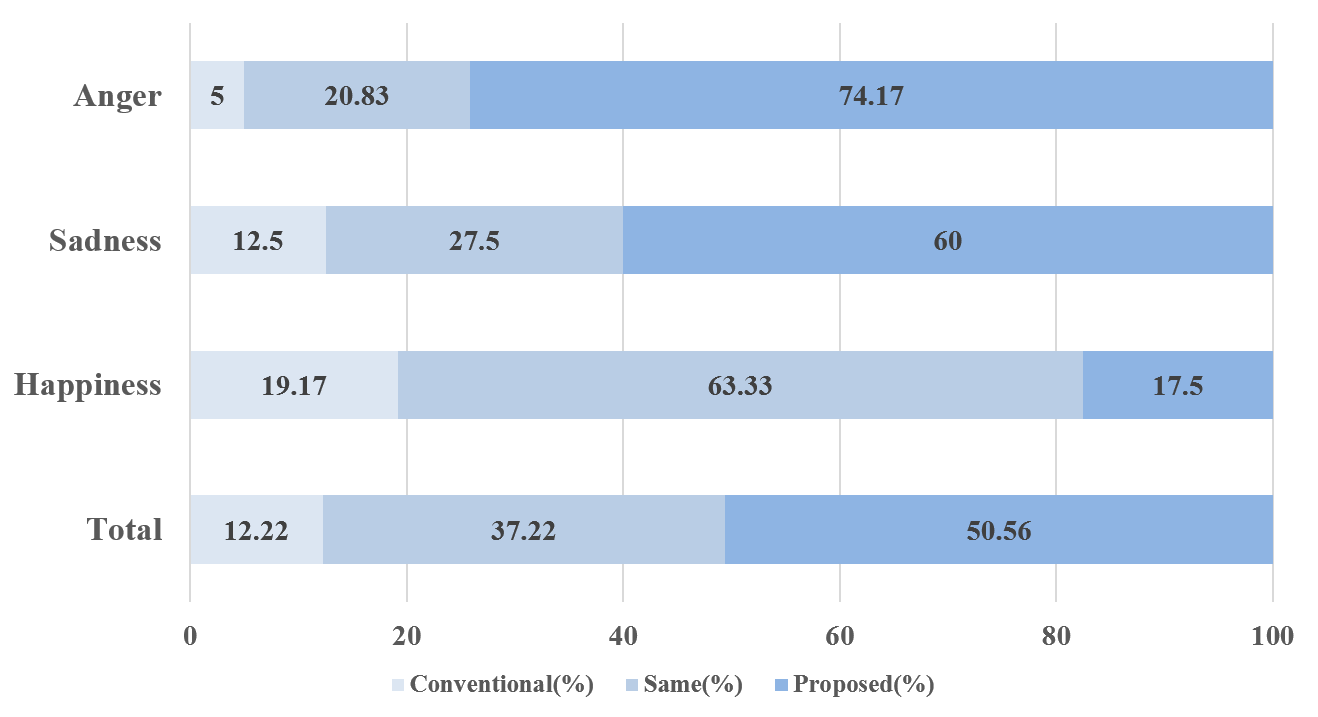}
    \caption{Preference of clarity about emotion expression}
    \label{fig:exp1}
\end{figure}
\section{Experiments and results}
\label{sec:experiments}

In this section, we evaluate the performance of our two proposed methods in terms of i) how clearlys the emotion in the synthesized speech signal is expressed, and ii) how to control the intensity of emotion from neutral to target emotions. Twelve people participated in the experiments.  

\subsection{Training setup}
We used the GST-Tacotron network as a baseline emotional TTS system \cite{wang2018style}, and trained it with a Korean male voice database. To  improve the TTS network performance, Tacotron was pre-trained with a large amount of neutral data. It consists of 6798 utterances and takes 17.56 hours. Then, Tacotron and WaveNet are trained to synthesize emotional speech signals.
After manually classifying a huge set of drama scripts into four emotion categories, such as happiness, sadness, anger, and neutral, we recorded approximately 740 utterances per emotion category. The total number of utterance was 2,956, which corresponded to 3.79 hours. Audio samples were sampled at a 16 kHz sampling rate and were quantized to 16 bits per sample.

\subsection{Emotion expression}
\label{exp1-1}

We first performed a preference test to compare the emotional expressiveness of the proposed I2I with the conventional mean-based method. Participants were asked to select the one that represents emotion clearly. If there was no detectable difference between them, they should choose the same. We used 10 unseen sentences for each emotion in the experiment. Figure \ref{fig:exp1} shows that the proposed I2I method outperforms the conventional method when synthesizing speech, except for the happiness category. Because the happiness emotion signal was recorded to have similar prosody to the neutral signal, the difference between two methods was not dramatic. The results confirm that the proposed method effectively extracts representative weight vectors for each emotion category compared to the mean-based approach. 

\subsection{Quality of synthesized speech signal}
\label{exp1-2}
To measure the quality of synthesized speech, we conducted mean opinion score (MOS) tests between the conventional mean-based and the proposed I2I-based methods. Subjects were asked to give a score based on the following standard: 1 = Bad; 2 = Poor; 3 = Fair; 4 = Good; and 5 = Excellent. The experimental settings were the same as the previous experiment. Results given in Table \ref{tb:exp1-2} verify that there is no significant difference between the two methods. Because a p-value computed by Welch's t-test \cite{welch1947generalization} which takes the null hypothesis that both groups have equal means is bigger than 0.05 for all emotion categories, null-hypothesis is retained and two data set have statistical insignificance. Unlike the previous experiment described in Section \ref{exp1-1}, anger has a slightly lower score. This is because people feel that a word with a strong emotion due to high pitches or stress feels a bit noisy. However, the proposed approach still generates high quality synthesized speech. From two experiments described in Sections \ref{exp1-1} and Section \ref{exp1-2}, the proposed I2I method meets the necessary qualifications for representative weight vectors mentioned in Section \ref{concept}. 

\begin{table}[t]
\caption{MOS of synthesized speech signal quality with 95\% confidence intervals.}\label{tb:exp1-2}
\begin{center}
\begin{tabular}{ccccc}
\Xhline{3\arrayrulewidth}
Emotion &Conventional&Proposed&p-value \\
\hline\hline
Anger        &3.91$\pm$0.53   &3.66$\pm$0.60 &0.36 \\
\hline
Happiness     &4.22$\pm$0.41        &4.18$\pm$0.45 &0.70 \\
\hline
Sadness       &3.99$\pm$0.51      &3.88$\pm$0.50 &0.53\\
\Xhline{3\arrayrulewidth}
Total &4.04$\pm$0.49    &3.90$\pm$0.54   \\
\Xhline{3\arrayrulewidth}
\end{tabular}
\end{center}
\end{table}

\subsection{Granularity of emotional expressiveness}
\label{exp2}
To evaluate the capability of emotional granularity, we gradually change the interpolation weight between the neutral and target emotions, i.e., anger, happiness, and sadness. We chose 10 unseen sentences and synthesized each sentence with six different weight ratios (linear ratio:$\{0, 0.2, 0.4, 0.6, 0.8, 1\}$ and SA-I2I one). Subjects were asked to choose the stronger emotion sample between two samples that are synthesized by the neighboring interpolation ratio pairs, e.g. (0, 0.2), (0.2, 0.4), and so on. For example, when samples A and B are synthesized at ratios of 0.2 and 0.4, respectively, selecting A is 'False' and selecting another is 'True'. If the emotional intensity of two samples was similar, they were asked to choose 'Same'.  

\begin{table}[t]
\caption{Recognition accuracy between neighboring emotion intensity. As it goes from \nth{1} to \nth{5}, the intensity of emotions increases.}\label{tb:exp2}
\begin{center}                                                                                                                          
\begin{tabular}{c|c|c|c|c|c|c}
\Xhline{3\arrayrulewidth}
Emotion &Method &\nth{1}&\nth{2}&\nth{3}&\nth{4}&\nth{5} \\
\hline\hline
  \multirow{2}{*}{Anger} &Linear &91.7 &90.0 &85.8 &81.7 &60.0 \\ 
  & SA-I2I &96.7 &93.3 &88.3 &80.0 &72.5 \\ \cline{1-7}
  \multirow{2}{*}{Happiness} &Linear &85.0 &55.0 &53.3 &43.3 &53.3 \\ \
  & SA-I2I &85.8 &44.2 &64.2 &52.5 &70.0 \\ \cline{1-7}
  \multirow{2}{*}{Sadness} &Linear &69.2 &88.3 &85.8 &62.5 &50.0 \\
  & SA-I2I &84.2 &87.5 &70.0 &82.5 &54.2 \\ \cline{2-7}
\Xhline{3\arrayrulewidth}
\end{tabular}
\end{center}
\end{table}

As shown in Table \ref{tb:exp2}, people perceived that they could successfully identify the intensity difference in most synthesized signals, especially anger and sadness. Compared to the linear interpolation-based approach, the SA-I2I method shows much higher accuracy, except for a few cases.  
Table \ref{tb:exp2} proves that the SA-I2I method, which extracts representative values used in synthesized speech considering the distribution of emotions, can express emotions more efficient than the linear interpolation method. 

\section{conclusion}
\label{sec:conclusion}
In this paper, we proposed a method to select representative weight vectors effectively to provide a rich emotional expression for a synthesized speech signal. To improve the clarity of emotional expressiveness futher, we proposed an inter-to-intra emotion ratio-based approach that considered embedding distances between inter and intra-categorical style token weights.
In addition, we also applied an effective interpolation method to change the intensity of emotional expressiveness flexibly. We first generated weight vectors to have a new distribution depending on the emotion intensity, then we found representative vectors using the proposed I2I method. The experiments demonstrated the superiority of the proposed approach over the conventional mean-based approach. Our future work is to apply it to a multi-speaker environment to produce unified style embeddings applicable for all people, regardless of age, gender, etc.

\section{acknowledgement}
\label{sec:acknowledge}
This work was supported by Institute for Information \& communications Technology Promotion (IITP) grant funded by the Korea government (MSIP). (2015-0-00860, Development of assistive broadcasting technology for invisible and deaf people's media accessibility)

\vfill\pagebreak
\bibliographystyle{IEEEbib}
\bibliography{refs}

\end{document}